\documentclass[article,prl,balancelastpage,twocolumn,showpacs]{revtex4}
\usepackage{amsfonts}
\usepackage{makeidx}
\usepackage{amssymb}
\usepackage{dcolumn}
\usepackage{graphicx}
\usepackage{acronym}

\input{tcilatex}

\begin{document}

\title{Energy momentum tensor in the nonsymmetric gravity}
\author{B. V. Gisin }
\affiliation{IPO, Ha-Tannaim St. 9, Tel-Aviv 69209, Israel. E-mail:
borisg2011@bezeqint.net}
\date{}

\begin{abstract}
General relativity is the theory with unclear energy momentum tensor. The
proposed approach allows to construct the energy momentum tensor for
relativity with non-symmetric metric. A consequence of the approach is
confirmed in the nuclear physics and may be applicable to astrophysics
\end{abstract}

\pacs{04.20.Cv, 04.20.Ex, 04.50.Kd, 04.40.Nr \hspace{20cm}}
\maketitle

\section{Introduction}

The theory of relativity allows the possibility of a nonsymmetric metric
tensor. A. Einstein, in his attempts to build out an Unified Field Theory,
associated the antisymmetric part of this tensor with electromagnetism \cite%
{Ein}.

Afterwards it was found that the antisymmetric metric tensor may represent a
generalized gravity \cite{Moff} with a new forces and properties \cite{Ham}, 
\cite{Pop}.

Presently, Einstein's interpretation is not popular among the physical
community. This scepticism is expressed in the phrase: "Research in this
direction ultimately proved fruitless; the desired unified field theory was
not found". One of the problems is the transition to classical
electrodynamics. Another problem is the energy momentum tensor.

However, the potential of Einstein's interpretation is not exhausted.

The paper\ presents an approach, which on the one hand considers Einstein's
interpretation, on the other hand is applicable to the generalized Moffat
gravity. The distinguish feature of the approach is the construction of the
energy momentum tensor so that the field equation are fulfilled due to
Maxwell's equation.

In this way, in particular, the result obtained, that the current density
squared corresponds to a mass density.\ The paper demonstrates that this
surprising result is confirmed by experimental data.

\section{Field equations}

\subsection{The approach}

Usually the field equations are governed by the principle of least action 
\cite{Ein}, \cite{W}. The Lagrangian is formed from the metric tensor and
the Ricci tensor. The main argument in favor of the variational principle is
the compatibility of equations in the system. In this case the energy
momentum tensor (for brevity the energy tensor) must be postulated: "Our
problem now is introduce a tensor $T_{\mu \nu }$, of the second rank, whose
structure we do know but provisionally\ldots " \cite{Ein}.

In the present approach this argument has no decisive significance. The
energy tensor is constructed by inserting new terms so that the field
equation turns into equality due to Maxwell's equation. Such a construction
may consist of the inclusion of missing terms as well as the compensation of
redundant terms. All inserts in the tensor should be material, that is, they
should disappear when the electric current goes to a zero.

In a sense, the structure of Maxwell's equation is optimal for any field
described by an antisymmetric tensor of the second rank. Therefore, for
simplicity, we use here the terminology of electrodynamics, in spite of the
fact that this approach is applicable also to Moffat's interpretation.

In our approach we come back to the initial Einstein's works. We postulate
that the covariant derivative of the metric tensor equals zero, and equate
the Einstein tensor to energy tensor.

Generally the symmetric $g_{(\mu \nu )}$ and antisymmetric $g_{<\mu \nu >}$
part of the metric tensor describes the gravitational and electromagnetic
field. However, for a better understanding of the role of the
electromagnetic field and the principle of constructing the energy tensor,
preferably to consider the case of absence of the gravitational field. In
this case $g_{(\mu \nu )}=\delta _{\mu \nu }$ and the interval has the
Euclidean form$.$ However all manipulations with such a form of the metric
tensor allows only the orthogonal coordinate transformations. Therefore we
assume a very weak gravitational field, $\gamma _{\mu \nu }=g_{(\mu \nu
)}-\delta _{\mu \nu },$ $|\gamma _{\mu \nu }|\ll |g_{<\mu \nu >}|$, which
allows more general transformations. In addition, we assume also $|g_{<\mu
\nu >}|\ll 1.$

Using the assumptions, we expand parameters of the field in power series in
the electromagnetic tensor. This leads to arising symmetric terms in the
energy tensor. The terms may be sources of the gravitational field. The
field has the order of the electromagnetic field squared.

For simplicity, we use the expansion including the second order. It is
enough to demonstrate main features of the approach.

The approach provides a simple algorithm of matching the field equation and
Maxwell's equation for every order of the approximation.

\subsection{The metric tensor and Christoffel's symbol}

We use the normalized (dimensionless) metric tensor, electric current
density and coordinates with $x_{4}=it$.

The covariant derivative of vectors $V_{\mu }$ and $V^{\mu }$ is defined as%
\begin{equation}
V_{\mu ;\nu }=V_{\mu ,\nu }-V_{\sigma }\Gamma _{\mu \nu }^{\sigma },\text{ \ 
}V_{;\nu }^{\mu }=V_{,\nu }^{\mu }+V^{\sigma }\Gamma _{\sigma \nu }^{\mu },
\label{AD}
\end{equation}%
the semicolon and comma denotes the covariant and partial differentiation,
respectively.

We assume that

\begin{equation}
g_{\mu \nu }=\delta _{\mu \nu }+\varphi _{\mu \nu }+\gamma _{\mu \nu }.
\label{g0}
\end{equation}%
where $\delta _{\mu \nu }$ is the Kronecker delta, $\varphi _{\mu \nu }$ is
the electromagnetic tensor.

The covariant derivative of the metric tensor obeys the equation%
\begin{equation}
g_{\mu \nu ;\sigma }\equiv g_{\mu \nu ,\sigma }-g_{\kappa \nu }\Gamma _{\mu
\sigma }^{\kappa }-g_{\mu \kappa }\Gamma _{\sigma \nu }^{\kappa }=0,
\label{gh}
\end{equation}%
Both the tensors $g^{\mu \nu }$and $g_{\mu \nu }$ are connected by the
relation%
\begin{equation}
g^{\sigma \mu }g_{\sigma \nu }=\delta _{\nu }^{\mu }=g^{\mu \sigma }g_{\nu
\sigma }  \label{ghl}
\end{equation}%
From this relation $g^{\mu \nu }$ is defined as the corresponding minor
divided by the determinant $g=$ $|g_{\mu \nu }|.$

The Christoffel symbol $\Gamma _{\mu \nu }^{\sigma }$ can be expressed as a
function of the metric tensor $g_{\mu \nu },$ using the algebraic equation (%
\ref{gh}) relative $\Gamma _{\mu \sigma }^{\kappa }$ and $\Gamma _{\sigma
\nu }^{\kappa }$. However, in contrast to the symmetric case, this
expression is too sophisticated. Therefore we use successive approximations
expanding $\Gamma _{\mu \nu }^{\sigma }$ in power series in $\varphi _{\mu
\nu }$ and taken into account that $\gamma _{\mu \nu }$ can be presented as
a sum of even power in $\varphi $.

With help of Eq. (\ref{g0}) express $\varphi _{\nu \mu ;\sigma }=(g_{\mu \nu
}-g_{\nu \mu })_{;\sigma }$ and $(g_{\mu \nu }+g_{\nu \mu })_{;\sigma }$ in
terms of the metric tensor and Christoffel's' symbols. Then substitute the
expression (\ref{g0}) and make the cyclic interchange $\mu ,\nu ,\sigma ,$
we obtain the expansion, including the terms of the second order%
\begin{eqnarray}
\Gamma _{\mu \nu }^{\sigma } &=&\varphi _{\mu \nu ,\sigma }+\varphi _{\kappa
\nu }\varphi _{\mu \sigma ,\kappa }+\varphi _{\mu \kappa }\varphi _{\sigma
\nu ,\kappa }+  \nonumber \\
&&+\frac{1}{2}(\gamma _{\sigma \mu ,\nu }+\gamma _{\nu \sigma ,\mu }-\gamma
_{\mu \nu ,\sigma })+\ldots ,  \label{g2}
\end{eqnarray}

\subsection{The Maxwell equation}

In the Euclidean space the equation can be written as follows 
\begin{equation}
\varphi _{\{\mu \nu ,\sigma \}}=0,\text{ }\varphi _{\mu \nu ,\nu }=j_{\mu },
\label{M0}
\end{equation}%
where the bracers mean the cyclic interchange\ $\mu ,\nu ,\sigma $. $j_{\mu
} $ is the current density%
\begin{equation}
\text{\ }j^{\mu }\text{ }=\rho \frac{dx^{\mu }}{ds},  \label{jcon}
\end{equation}%
$\rho $ is the charge density. In the Euclidean space co- and -contravariant
vectors coincides.

It is well know from quantum theory that the potential is a more fundamental
characteristic than components of the electromagnetic field. In the
Euclidean space $\varphi _{\mu \nu }$ can be expressed in terms of the
potential as follows 
\begin{equation}
\varphi _{\mu \nu }=A_{\mu ,\nu }-A_{\nu ,\mu }  \label{pot}
\end{equation}

Due to the expression (\ref{pot}) the first equation in (\ref{M0}) is
fulfilled identically. But the Einstein tensor contain $\varphi _{\mu \nu }$
with two derivatives. Keeping it in mind, differentiate the first equation
of (\ref{M0}) in respect to $x_{\sigma }$. Then, using the second equation,
obtain%
\begin{equation}
\varphi _{\mu \nu ,\sigma \sigma }-j_{\mu ,\nu }+j_{\nu ,\mu }=0.  \label{Ms}
\end{equation}%
This equation integrates both the field $\varphi _{\mu \nu }$ and matter $%
j_{\mu }$. We should expect appearance of this equation in the first
approximation of the field equation \cite{eta}

\subsection{The Einstein tensor and field equation}

Einstein's tensor is equivalent to that in the general relatively 
\begin{equation}
G_{\mu \nu }=R_{\mu \nu }-\frac{1}{2}g_{\mu \nu }R.  \label{G}
\end{equation}%
where $R_{\mu \nu }$ is the Ricci tensor. 
\begin{equation}
R_{\mu \nu }=\Gamma _{\mu \nu ,\rho }^{\rho }-\Gamma _{\mu \rho ,\nu }^{\rho
}+\Gamma _{\mu \nu }^{\kappa }\Gamma _{\kappa \rho }^{\rho }-\Gamma _{\mu
\rho }^{\kappa }\Gamma _{\kappa \nu }^{\rho }.  \label{R}
\end{equation}

The covariant derivative of $G_{\mu \nu }$ equals zero. Equating $G_{\mu \nu
}$ to the energy tensor $G_{\mu \nu }=T_{\mu \nu }$ we obtain the field
equation similarly to the symmetric theory. For convenience, the modified
energy tensor $T_{\mu \nu }^{\ast }=T_{\mu \nu }-$ $\frac{1}{2}g_{\mu \nu
}T, $ where $T=g^{\mu \nu }T_{\mu \nu },$\ is in common use. Thus the field
equation is as follows%
\[
R_{\mu \nu }=T_{\mu \nu }^{\ast }. 
\]

\subsubsection{The first approximation}

In the first approximation 
\begin{equation}
\Gamma _{\mu \nu }^{\sigma }=\varphi _{\mu \nu ,\sigma },\text{ \ }R_{\mu
\nu }=\varphi _{\mu \nu ,\sigma \sigma }-\varphi _{\mu \sigma ,\sigma \nu },
\label{R1}
\end{equation}%
we find the field equation 
\begin{equation}
\varphi _{\mu \nu ,\sigma \sigma }-\varphi _{\mu \sigma ,\sigma \nu }=T_{\mu
\nu }^{\ast }.  \label{fq1}
\end{equation}

A comparison with (\ref{Ms}), in accordance with our approach, allows to
construct the energy tensor in the first approximation 
\begin{equation}
T_{\mu \nu }^{\ast }=-j_{\nu ,\mu },\text{ \ }  \label{T1}
\end{equation}%
This is an example of the inserted term.

Eq. (\ref{fq1}) easily can be rearranged as follows%
\begin{equation}
(\varphi _{\mu \nu ,\sigma }+\varphi _{\sigma \mu ,\nu }+\varphi _{\nu
\sigma ,\mu })_{,\sigma }-\varphi _{\nu \sigma ,\mu \sigma }=-j_{\nu ,\mu }
\label{fp}
\end{equation}%
Provided the condition (\ref{pot}), we obtain $\varphi _{\nu \sigma ,\sigma
}=j_{\nu }$ and the Maxwell equation (\ref{M0}).\ 

$j_{\nu ,\mu }$ is a 'strange' term since corresponding energy density
changes the sign together with electromagnetic field. Moreover its
contribution in energy $\int j_{4,4}dv$ vanishes by the integration over all
the spatial volume, in both the conditions: $j_{\mu }=0$ at the boundary of
integration and the current conservation \ $j_{\mu ,\mu }=0$.

\subsubsection{The second approximation}

Using Eqs. (\ref{g2}), (\ref{M0}), it can be shown that the antisymmetric
part of the Ricci tensor equals 
\begin{equation}
R_{<\mu \nu >}=\varphi _{\mu \nu ,\sigma }j_{\sigma }.  \label{Ra2}
\end{equation}%
Obviously, the contribution in energy from this term also equal zero since%
\emph{\ }$\varphi _{44}\equiv 0$\emph{.} Accordingly to our approach, this
term in the covariant form should be inserted in the energy tensor 
\begin{equation}
T_{\mu \nu }=-j_{\nu ;\mu }+2j^{\sigma }\varphi _{\mu \nu ;\sigma }.
\label{Tc}
\end{equation}%
This is an example of the compensating term.\emph{\ }

It is noteworthy that the term $j_{\nu ;\mu }$ in the second approximation
exactly coincides with the term $j^{\alpha }\varphi _{\mu \nu ;\sigma }$ in
the first approximation. Therefore the inserted term in $T_{\mu \nu }$ is
doubled.

The symmetric part of the Ricci tensor in the second approximation is as
follows%
\[
R_{(\mu \nu )}=j_{\mu }j_{\nu }+(\varphi _{\mu \kappa }\varphi _{\sigma \nu
})_{,\kappa \sigma }-\frac{1}{4}\varphi _{\sigma \kappa ,\mu \nu }^{2}+ 
\]%
\begin{equation}
+\frac{1}{2}(\gamma _{\sigma \mu ,\nu \sigma }+\gamma _{\nu \sigma ,\mu
\sigma }-\gamma _{\mu \nu ,\sigma \sigma }-\gamma _{\sigma \sigma ,\mu \nu
}).  \label{Rs2}
\end{equation}

The term%
\begin{equation}
j^{\mu }j^{\nu }=\rho ^{2}\frac{dx^{\mu }}{ds}\frac{dx^{\nu }}{ds},
\label{jj}
\end{equation}%
is of special interest. Its shape is similar to the mass term in the general
relativity with symmetric metric, but the role of $\rho ^{2}$ plays the mass
density. Since we use the dimensionless units the mass density should be
proportional to the gravitational constant, which is determined by comparing
the field equation with the Newtonian theory of gravity.

From Eq. (\ref{jj}) it follows that the charge density squared $\rho ^{2}$
corresponds to a mass density \cite{Gis}. This parameter, analogously to the
mass density, is proportional to an electrodynamic constant. For this
constant only the dimension is known.

The term $j_{\mu }j_{\nu }$ must be inserted in the energy tensor.
Conveniently to divide the inserts into two parts. The first part is the
compensating term (\ref{jj}), which inserted uniquely. The second part is
inserted\ as the same term but with a constant $C$ on the same basis as in
the general theory of relativity.

With this definitions $T_{\mu \nu }^{\ast }$ takes the form \ \ 
\begin{equation}
T_{\mu \nu }^{\ast }=-j_{\nu ;\mu }+2j^{\alpha }\varphi _{\mu \nu ;\sigma
}+(1+C)j_{\mu }j_{\nu }.  \label{T2}
\end{equation}%
The gravitational field obeys field equation 
\begin{equation}
\frac{1}{2}(\gamma _{\sigma \mu ,\nu \sigma }+\gamma _{\nu \sigma ,\mu
\sigma }-\gamma _{\sigma \sigma ,\mu \nu }-\gamma _{\mu \nu ,\sigma \sigma
})+  \label{gmn}
\end{equation}%
\begin{equation}
+(\varphi _{\mu \kappa }\varphi _{\sigma \nu })_{,\kappa \sigma }-\frac{1}{4}%
\varphi _{\sigma \kappa ,\mu \nu }^{2}=Cj_{\mu }j_{\nu }.  \label{eqf}
\end{equation}

In the general~relativity a supplementary condition is imposed for
simplification of the field equation. The purpose of the simplification is
to cancel the first three terms in (\ref{gmn})%
\begin{equation}
\gamma _{\sigma \mu ,\nu \sigma }+\gamma _{\nu \sigma ,\mu \sigma }-\gamma
_{\sigma \sigma ,\mu \nu }=0.  \label{g3}
\end{equation}%
This condition consists of 16 equations connecting 10 functions $\gamma
_{\mu \nu }$, some of the equations coincide. However the number of
equations may be reduced by symmetry. The same result can be obtained using
the four conditions\emph{\ }$\Xi _{\mu \nu ,\nu }=0$,\ where%
\begin{equation}
\Xi _{\mu \nu }=\gamma _{\mu \nu }-\frac{1}{2}\delta _{\mu \nu }\gamma
_{\sigma \sigma }.  \label{Xi}
\end{equation}%
The four conditions are interpreted as an "analog of the Lorentz condition"
in electrodynamics.\emph{\ }

Under this condition the equation for the gravitational field is simplified%
\begin{equation}
-\frac{1}{2}\gamma _{\mu \nu ,\sigma \sigma }+(\varphi _{\mu \kappa }\varphi
_{\sigma \nu })_{,\kappa \sigma }-\frac{1}{4}\varphi _{\sigma \kappa ,\mu
\nu }^{2}=C\rho ^{2}\frac{dx^{\mu }}{ds}\frac{dx^{\nu }}{ds}.  \label{Eq}
\end{equation}

Up to now we considered the gravitational and electromagnetic field as
originated from one source.

Assume, that the source of a gravitation field (not connected with an
electromagnetic field) exists. We divide the total mass of any charged
object as well as nuclei\ or particles on the proper mass and mass connected
with the charge.

In that case instead the factor $C$ in right part of Eq. (\ref{eqf}) would
be the factor $(\gamma +C\rho ^{2})$ and the tensor $\gamma _{\mu \nu }$ is
the tensor of total gravitational field originated from both the mass
density $\gamma $ and the mass density connected with the charge density $%
C\rho ^{2}$.

A symmetric part of the energy tensor in that case, exclusive of \ $(-j_{\nu
;\mu }),$ would be $T_{\mu \nu }=T_{\mu \nu }^{\ast }-\frac{1}{2}g_{\mu \nu
}T^{\ast }$%
\begin{equation}
T_{\mu \nu }=[\gamma +(1+C)\rho ^{2}]\frac{dx^{\mu }}{ds}\frac{dx^{\nu }}{ds}%
-.\frac{1}{2}g_{\mu \nu }(1+C)\rho ^{2}.  \label{Ted}
\end{equation}

In the approximation of low velocities and weak fields 
\begin{equation}
T_{44}\approx \gamma +\frac{1}{2}(1+C)\rho ^{2}  \label{md}
\end{equation}%
is the total mass density.

Surprisingly the last term in the expression (\ref{md}) can be found in the
semi-empirical mass formula.

\subsection{The Bethe--Weizs\"{a}cker mass formula}

The semi-empirical mass formula is used to approximate the mass of an atomic
nucleus

\begin{equation}
m=Zm_{p}+Nm_{n}-a\frac{(A-2Z)^{2}}{A}+E_{b}(A,Z),  \label{BW}
\end{equation}%
where $Z$ and $N$ is the number of protons and neutrons, $A=$ $Z+N$ is the
total number of nucleons, $m_{p}$ and $m_{n}$ are the rest mass of a proton
and a neutron, respectively, $E_{b}$ the binding energy of the nucleus, $a$
and $E_{b}$ are small and determined empirically.

The mass corresponding to $\rho ^{2}$ is defined by the integral over the
volume of nucleus. In the liquid drop model of nucleus approximately 
\begin{equation}
\int \rho ^{2}dv\approx \frac{Z^{2}}{A}.  \label{Cint}
\end{equation}%
Eq. (\ref{BW}) can be changed to following form 
\begin{equation}
m=Zm_{pe}+Nm_{ne}-4a\frac{Z^{2}}{A}+E_{b},  \label{BWr}
\end{equation}%
where $m_{pe}$ and \ $m_{ne}$ is the effective mass of the proton and
neutron in nucleus 
\begin{equation}
m_{pe}=m_{p}+3a,\ m_{ne}=m_{n}-a.  \label{me}
\end{equation}

The three first terms in the right part of Eq. (\ref{BWr}) correspond to the
integral of the term $T_{44}$ from Eq. (\ref{Ted}) over all the volume of
nucleus 
\begin{equation}
Zm_{pe}+Nm_{ne}-4a\frac{Z^{2}}{A}\rightarrow \int [\gamma +\frac{1}{2}%
(1+C)\rho ^{2}]dv.  \label{mel}
\end{equation}

Accordingly to Eq. (\ref{mel}), $C<0,$ moreover\textit{\ }$(1+C)<0.$
Nevertheless, for any atomic nucleus the integral in (\ref{mel}) always is
positive.

However, if the charge density increases and becomes larger than nuclear,
gravity may be replaced by repulsion and energy may change the sign.

\section{Conclusion}

We have considered the general relativity with nonsymmetric metric and
proposed the approach for the construction of the energy momentum tensor.

It was shown that the energy tensor can be build out from the inserted
material terms so that the field equation turns into the equality due to
Maxwell's equation.

Distinctive feature of \ the approach is the start from the first
approximation without the gravitational field. In the second approximation a
source of gravitational field arises. This source is similar to that in the
general relativity but with the mass density proportional to the charge
density squared.

The term corresponding to the source finds surprising confirmation in
nuclear physic. Another surprise is the negative sing of this mass density.

If such term is really applicable to nuclear physics then increasing the
charge density may have far-reaching consequences from microphysics to
astrophysics. However, for that we should use exact expression for the
Christoffel symbols as a function of the metric tensor.

In astrophysics the charge density, which corresponds to the zeroth energy
density, would determine the beginning of the Big Bang.


\begin{thebibliography}{9}
\bibitem{eta} In \cite{Ein} the electromagnetic field $\frac{1}{2}\eta _{\mu
\nu \sigma \rho }\varphi ^{\sigma \rho }$ is used, where $\eta _{\mu \nu
\sigma \rho }$ is the Levi-Chivita tensor. This tensor interchanges electric
and magnetic field.
\end{thebibliography}
\end{document}